\documentclass[twocolumn,prd,superscriptaddress,amsmath,amssymb]{revtex4}
\usepackage{graphicx,hyperref}
\newcommand{\slp}{ {p\mspace{-7.5mu}/\mspace{-1.5mu}} }
\newcommand{\e}{\epsilon}
\newcommand{\img}{\mathrm{i}}
\newcommand{\df}{\mathrm{d}}
\newcommand{\w}{\omega}
\newcommand{\D}{\Delta}
\newcommand{\cD}{\mathcal{D}}
\newcommand{\cO}{\mathcal{O}}
\newcommand{\cP}{\mathcal{P}}
\newcommand{\ord}{\mathcal{O}}
\newcommand{\vevB}[1]{\langle #1 \rangle_{\! B}}
\newcommand{\hQ}{\hat{Q}}
\newcommand{\hp}{\hat{p}}
\newcommand{\hep}{\hat{\epsilon}}
\newcommand{\hk}{\hat{k}}
\newcommand{\hl}{\hat{l}}
\newcommand{\bn}{\bar{n}}
\newcommand{\hla}{\hat{\lambda}}
\newcommand{\hrh}{\hat{\rho}}
\newcommand{\hta}{\hat{\tau}}

\begin{document}

\preprint{SI-HEP-2004-08}
\preprint{hep-ph/0408273}

\date{August 25, 2004}

\title{\boldmath Shape Function Effects in $B \to X_c \ell \bar{\nu}_\ell$}

\author{Thomas Mannel}
\affiliation{Theoretische Physik I,
Universit\"at Siegen,  D--57068  Siegen, Germany}
\author{Frank J. Tackmann}
\altaffiliation[Address after Sep 1, 2004: ]{Department of Physics,
University of California, Berkeley, CA 94720, USA}
\affiliation{Institut f\"ur Kern-- und Teilchenphysik,
Technische Universit\"at Dresden, D--01062 Dresden, Germany}
\begin{abstract}
Owing to the fact that
$m_c^2 \sim m_b \Lambda_\text{QCD}$, the endpoint
region of the charged lepton energy spectrum in the inclusive decay
$B \to X_c \ell \bar{\nu}_\ell$ is affected by the Fermi motion of the
initial-state $b$ quark bound in the $B$ meson. This effect is described
in QCD by shape functions. Including the mass of the final-state quark,
we find that a different set of operators as employed in Ref.~\cite{Bauer:2002yu}
is needed for a consistent matching, when incorporating the subleading contributions in
$B \to X_q \ell \bar{\nu}_\ell$ for both $q = u$ and $q = c$.
In addition, we modify the usual twist expansion in such a way that it yields
a description of the lepton energy spectrum which is not just valid in
the endpoint region, but over the entire phase space.
\end{abstract}


\maketitle

\section{Introduction}

The theoretical machinery for the determination of $\lvert V_{cb} \rvert$
from semileptonic $B$-meson decays has reached
a mature state over the last years. With the very precise data on
the  exclusive $B \to D^{(*)} \ell \bar{\nu}_\ell$ channels as well
as on the inclusive decays $B \to X_c \ell \bar{\nu}_\ell$ from the
$B$-meson factories the theoretical description has been improved so much
that currently relative theoretical uncertainties for $\lvert V_{cb} \rvert$ of less than
2\% are quoted \cite{Benson:2003kp,Bauer:2004ve}, yielding a
total uncertainty of about 2\% \cite{Aubert:2004aw}.

The nonperturbative corrections to the inclusive decays are parametrically
of order $1/m_b^2$ and hence are expected to be smaller than in the exclusive
channels, where the corrections are of order $1/m_c^2$. However,
the inclusive rate depends on the $b$-quark mass $m_b$, which needs to be
determined in a suitable scheme as precisely as possible.

In addition, also the parameters of the heavy quark expansion are needed, such as
$\lambda_1$ and $\lambda_2$ at order $1/m_b^2$ and the corresponding
parameters appearing at higher orders. These parameters are obtained
experimentally by taking moments of the various inclusive distributions (such as
lepton energy spectra or hadronic invariant mass distributions), but
the higher moments become more and more sensitive to higher-order
corrections in $1/m_b$, since the leading contribution to the
$n$th moment is roughly of the order $1/m_b^n$.

When determining the heavy quark parameters from the
lepton energy spectrum, the higher moments become sensitive to the endpoint
region of the spectrum. Using the $1/m_b$ expansion the leading term is
the partonic rate and still a smooth function, but already the
first non-vanishing nonperturbative contribution exhibits an irregular
behavior which is unphysical. The situation is in fact very similar to
the one in $b \to u$ transitions, where it is known that these
singular contributions can be resummed into a shape function.
For heavy to light decays, such as the decays $b \to u \ell \bar{\nu}_\ell$
and $b \to s \gamma$,
the twist expansion, that is, the resummation of nonperturbative contributions,
has been performed to the subleading level in the $1/m_b$ expansion
\cite{BLM1,Bauer:2002yu,BLM21,Burrell:2003cf}.

It has already been noticed some time ago \cite{Mannel:1994pm} that the light-cone distribution
of the $B$ meson also has a significant effect on the endpoint region of
the lepton energy spectrum in  $B \to X_c \ell \bar{\nu}_\ell$. This is
due to the fact that numerically $m_c^2 \sim m_b \Lambda_\text{QCD}$.
The charm quark mass thus has to be counted as $\sqrt{m_b \Lambda_\text{QCD}}$ when performing
the power counting. The endpoint region is known to be determined by the
light-cone distribution of the initial state and has the width
$\sqrt{ m_b \Lambda_\text{QCD}}$, which happens to be of the same order as
$m_c$. Therefore, it is useful to consider the effects of the light-cone
distribution of the $B$ meson also in $B \to X_c \ell \bar{\nu}_\ell$.

In the present paper we perform this analysis and compare with the standard
expansion. As a by-product, we suggest a modified twist expansion which can be applied
over the full phase space, incorporating the twist expansion in the endpoint region
as well as the usual local expansion in the rest of phase space.

Furthermore, performing the limit $m_c \to 0$ we discover an inconsistency
in comparison with previous work \cite{Bauer:2002yu}. It turns
out that at subleading order additional operators are needed which are
formally of leading order, but have coefficients of subleading order.
Expanding into the usual local expansion we
obtain the correct result for the terms of order $1/m_b^3$, indicating that our
result is consistent. As a consequence, compared to Ref.~\cite{Bauer:2002yu}
additional nonperturbative input in the form of new shape functions
appears.

\subsection{Inclusive Decay Rate}

Neglecting the masses of the leptons the energy spectrum
of the charged lepton in the $B$ rest frame is given via the optical
theorem by
\begin{equation}
\label{dGqdy}
\frac{\df\Gamma_q}{\df y} = 4 \Gamma_0 y^2 \theta(y) \vevB{T_q}
,\end{equation}
where $y = 2E_\ell/m_b$ denotes the rescaled lepton energy,
\begin{equation*}
\Gamma_0 = \frac{G_F^2 \lvert V_{qb} \rvert^2 m_b^5}{192\pi^3}
,\end{equation*}
and $q$ stands for either $u$ or $c$. The ``$B$ expectation value'' is defined as
$\vevB{\cO} = \langle B \lvert \cO \rvert B \rangle / 2m_B$, where the
QCD states are normalized to $2 m_B$.
The  operator $T_q$ has the form
\begin{equation}
\label{T_def}
T_q = \frac{48\pi^2}{m_b^3 y}
\mathrm{Im}\, \frac{\img}{\pi} \sum_{s_\ell} \int\df^4x\, e^{-\img p_\ell\cdot x}
T \bigl[ W_q^\dagger(x) W_q(0) \bigr]
,\end{equation}
where $s_\ell$ and $p_\ell$ are the lepton spin and momentum.
The effective weak current is
\begin{equation*}
\label{W_def}
W_q = (\bar{q} \gamma_\alpha P_L b) (\bar{u}_\ell \gamma^\alpha P_L \nu_\ell)
= (\bar{q} \gamma_\alpha P_L \nu_\ell) (\bar{u}_\ell \gamma^\alpha P_L b)
,\end{equation*}
with $P_L = (1 - \gamma_5)/2$ and $u_\ell$ denoting the lepton spinor.
Plugging this into Eq.~\eqref{T_def} yields
\begin{subequations}
\begin{align}
T_q &= \frac{48\pi^2}{m_b^3 y}
\biggl( -\frac{1}{\pi} \mathrm{Im}\, \bar{b} L_{\alpha\beta} b \Pi_q^{\alpha\beta} \biggr)
\quad \text{with} \\
\label{Lab_def}
L_{\alpha\beta} &= \gamma_\alpha \slp_\ell \gamma_\beta P_L
,\\
\label{Piab_def}
\Pi_q^{\alpha\beta} &= -\img\!\int\df^4x e^{\img (p_b - p_\ell)\cdot x}
T \bigl[ (\bar{\nu}_\ell \gamma^\alpha P_L q)(x) (\bar{q} \gamma^\beta P_L \nu_\ell)(0) \bigr]
.\end{align}
\end{subequations}
Here, $L_{\alpha\beta}$ is the leptonic tensor and $\Pi_q^{\alpha\beta}$ represents
the inclusive $q$-quark-neutrino loop with the momentum transfer $p_b - p_\ell$.
The  $b$-quark momentum $p_b$ contains a large part $m_b v$,
where $v$ is the velocity of the $B$ meson. As usual,
assuming that $Q = m_b v - p_\ell$ sets a perturbative scale, large compared
to $k = p_b - m_b v \sim \ord(\Lambda_\text{QCD})$,
we may perform an OPE of the transition operator $T_q$ in powers of
$\Lambda \equiv \Lambda_\text{QCD}/m_b$
\cite{ManoharWise,Bigiinclusive1,Mannelinc}.

\subsection{Light-Cone Vectors and Power Counting}

Using appropriate powers of $m_b$ we may work with dimensionless
variables, which are denoted by a hat, for example $\hQ = Q/m_b = v - \hp_\ell$.
We also define $\D = 1 - y$ and $\rho = m_c^2/m_b^2$.

As already noted, the $c$-quark mass satisfies
$m_c^2 \sim m_b \Lambda_\text{QCD}$ and thus $\rho \sim \ord(\Lambda)$.
The kinematic endpoint in the local OPE is given by $\D = \rho$, the partonic
endpoint. The endpoint region of the lepton energy
spectrum is defined by $\D \sim \ord(\Lambda)$, and it is well known \cite{ManoharWise}
that in this region the local OPE breaks down.  However, it has been
shown that one may still perform a light-cone or twist expansion
\cite{NeubertShape,NeubertShape1, BigiMotion}.

To set up the light-cone expansion we use the
velocity $v$ and the lepton momentum $p_\ell$ to define a basis of
two light-cone vectors,
\begin{subequations}
\begin{equation}
v = \frac{1}{2} (n + \bn), \quad p_\ell = E_\ell \bn = \frac{m_b}{2} y \bn
,\end{equation}
satisfying $n^2 = \bn^2 = 0$ and $n \cdot \bn = 2$. The  metric is
decomposed accordingly
\begin{equation}
\eta^{\mu\nu}
= \frac{1}{2} n^\mu \bn^\nu + \frac{1}{2} \bn^\mu n^\nu + \eta_\perp^{\mu\nu}
,\end{equation}
and a generic four-momentum $p$ can be written as
\begin{equation}
p^\mu = \frac{1}{2} p_- n^\mu + \frac{1}{2} p_+ \bn^\mu + p_\perp^\mu
,\end{equation}
where we defined
\begin{equation}
p_+ = n \cdot p, \quad p_- = \bn \cdot p,
\quad p_\perp^\mu = \eta_\perp^{\mu\nu} p_\nu
.\end{equation}
\end{subequations}

The power counting
\begin{equation}
\label{usual-counting}
\hQ_- = 1, \quad \hQ_+ = \D \sim \ord(\Lambda), \quad \rho \sim \ord(\Lambda),
\quad \hk \sim \ord(\Lambda)
,\end{equation}
yields the standard twist expansion by expanding
in powers of $\Lambda$, taking also into account that $\rho \sim \ord(\Lambda)$.

Note that this power counting becomes wrong
for small lepton energies, since $\D$ becomes of $\ord(1)$.
In this case the spectrum is described by the usual local OPE. However,
as we shall discuss below, with a slight modification of the twist expansion
it is possible to describe the spectrum also for small lepton energies.

The relevant kinematic variable in the OPE is
$\hp = \hp_b - \hp_\ell = \hQ + \hk$, the light-cone components of which are
\begin{equation}
\label{hp_lc}
\hp_+ = \D + \hk_+, \quad \hp_- = 1 + \hk_-,
\quad \hp_\perp = \hk_\perp
.\end{equation}
Obviously all $\hk_+$ dependence occurs in the combination
$\D_+ = \D + \hk_+$.
In the local OPE the complete $\hk$ dependence is expanded. In particular,
this produces terms of the form $\hk_+/\D$, which become large, of
$\ord(1)$, near the
endpoint. The twist expansion avoids these terms, since only the $\hk_-$ and
$\hk_\perp$ dependences are expanded.

\begin{figure*}%
\centering
\includegraphics[width=0.7\columnwidth]{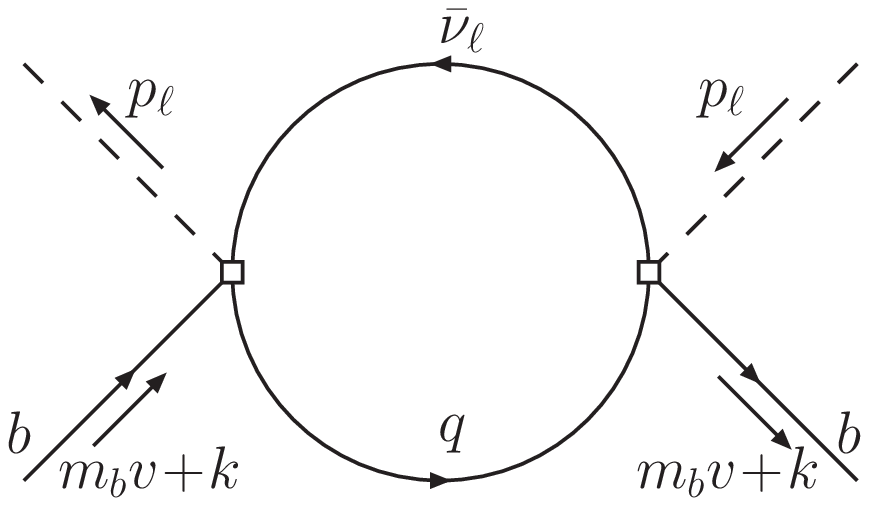}%
\hspace{\columnsep}
\includegraphics[width=0.7\columnwidth]{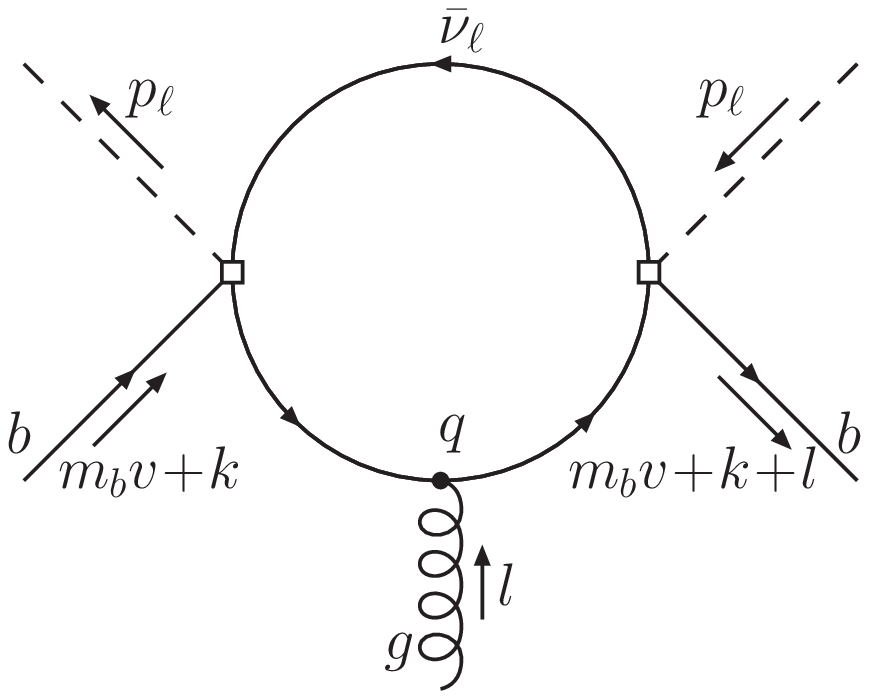}%
\caption{\label{fig:me}Feynman diagrams for the zero- and one-gluon matrix elements
(to leading order in $\alpha_s$).}
\end{figure*}

Alternatively to Eq.~\eqref{usual-counting}, we may as well
treat the complete dependence on $\D_+$ (and $\rho$) exactly. That is,
we use the power counting
\begin{equation}
\label{mod-counting}
\hk_- \sim \ord(\Lambda), \quad \hk_\perp \sim \ord(\Lambda)
,\end{equation}
and only expand in powers of $\hk_-$ and $\hk_\perp$ from the very beginning.
The order in $\Lambda$ of a twist term given by Eq.~\eqref{mod-counting} corresponds
to the order of the first local term it contains, once the usual local OPE is performed.
Eq.~\eqref{mod-counting} defines what we call the modified twist expansion and
holds for large \textit{and} small lepton energies.
Our results will therefore be valid over the whole lepton energy spectrum,
providing a correct interpolation between the usual twist and local expansions.

Our modified expansion is a direct extension of the usual twist expansion.
The latter does not expand $\Delta_+ = \Delta + \hk_+$ because both
$\Delta$ and $\hk_+$ are considered $\ord(\Lambda)$.
Once part of the $\hk_+$ dependence is left unexpanded, it is consistent
to keep the complete $\hk_+$ dependence unexpanded, as it just means
to keep correct small terms, which can now be treated exactly.
This is similar to the local expansion, where $\rho =m_c^2/m_b^2$, although
being numerically small, is always treated exactly, because it can be
treated exactly.
Once we exclude $\hk_+$ from the power counting and treat it exactly,
there is no need to count $\Delta$ as $\ord(\Lambda)$ anymore.
Thus we can treat it exactly much like $\rho$.
This in turn extends the validity of our expansion down to low
lepton energies (where numerically $\Delta$ is of order one which
caused the breakdown of the usual twist expansion).

In other words, starting from the full expression (including all
$\hk$-components) both expansions expand the $\hk_-$ and $\hk_\perp$
components. This is where we stop, while the usual expansion
additionally neglects terms of second and higher twist order
caused by $\Delta_+$, e.g., terms like $\Delta_+^2$. We keep all
those ``kinematic'' twist terms, because they are of $\ord(1)$ for low lepton
energies. Note that we do \textit{not} claim to include all second-order
twist contributions, i.e., our spectrum is only correct to $\ord(\Lambda)$
in the endpoint region, but it is correct to $\ord(\Lambda^2)$ for low
lepton energies.

Therefore, the difference to the usual twist expansion, defined by Eq.~\eqref{usual-counting},
is that the modified expansion automatically keeps all twist contributions that
in the usual power counting are of higher order only because of additional factors
of $\D_+$ or $\rho$. These contributions are purely kinematic and
do not require additional operators in the twist expansion, but appear only
as higher-order terms in the OPE coefficients. Taking them into
account yields a consistent result valid over the full region of lepton energies.

\section{Operator Product Expansion}

To keep things simple and to exhibit the structure of the OPE
we will perform it in terms of QCD light-cone operators.
Schematically, it has the form
\begin{equation}
T_q = \int\df\w \sum_n C^{(n)}(\w) \cO_n(\w)
.\end{equation}
In a second step the $B$ expectation values
of the $\cO_n(\w)$ are parametrized in terms of shape functions.

We expand to $\ord(\Lambda^2)$ as defined by Eq.~\eqref{mod-counting}.
With respect to the usual twist power counting this
includes all contributions of leading and subleading twist,
as well as some of second and even third order.
In other words, we obtain the full coefficients of all operators
with up to two covariant derivatives, which in particular
retains all local terms up to local $\ord(\Lambda^2)$.
To do so, we need to evaluate the zero- and one-gluon matrix elements
of $T_q$, depicted in Fig.~\ref{fig:me},
which we do to leading order in $\alpha_s$.

\subsection{Leading Twist}

We first consider the case $q = c$ and later take the limit
$m_c \to 0$.
The zero-gluon matrix element of $T_c$ yields the well-known result
\begin{multline}
\label{Tc_bb}
\langle b \lvert T_c \rvert b \rangle
= \theta(\hp^2 - \rho) \Bigl( 1 - \frac{\rho}{\hp^2} \Bigr)^2
\\ \times
\biggl[ \Bigl( 1 + 2\frac{\rho}{\hp^2} \Bigr) 3\eta_{(\mu\nu} \bn_{\alpha)} \hp^\mu \hp^\nu
- 3\rho \bn_{\alpha} \biggr] \bar{u}_b \gamma^\alpha P_L u_b
.\end{multline}
Indices in round brackets are completely symmetrized,
\begin{equation*}
\eta_{(\mu\nu} \bn_{\alpha)} = \frac{1}{3} ( \eta_{\mu\nu} \bn_\alpha
+ \eta_{\mu\alpha} \bn_\nu + \eta_{\alpha\nu} \bn_\mu )
.\end{equation*}
Extracting the terms of $\ord(1)$,
\begin{equation*}
\langle b \lvert T_c \rvert b \rangle
= \theta(\D_+ - \rho)
\Bigl(C^{(0)}_\alpha(-\hk_+)  
+ \ord(\Lambda) \Bigr) \bar{u}_b \gamma^\alpha P_L u_b
,\end{equation*}
we obtain the leading term in the OPE of $T_c$,
\begin{equation}
\label{Tc_to-0,0}
T_c = \int\df\w \theta(\D_\w - \rho)
\Bigl( C_\alpha^{(0)}(\w) \cO_0^\alpha(\w) + \ord(\Lambda) \Bigr)
,\end{equation}
where $\D_\w = \D - \w$. The leading operator has the form
\begin{equation}
\label{leading-op}
\cO_0^\alpha(\w) = \bar{b} \delta(\img \hat{\cD}_+ + \w) \gamma^\alpha P_L b
,\end{equation}
and its coefficient is
\begin{equation}
C^{(0)}_\alpha(\w)
= (1 - 3R_\w^2 + 2R_\w^3) n_\alpha + (2 - 3R_\w + R_\w^3)\D_\w  \bn_\alpha
,\end{equation}
with $\img\cD = \img D - m_b v$ and $R_\w = \rho/\D_\w$.
Note that in the modified twist expansion we keep the
contributions proportional to $\bn_\alpha$, which would usually
be considered as subleading twist due to the additional factor
of $\D_\w$.

The $B$ expectation value of $\cO_0^\alpha(\w)$ is given to leading order
by the leading shape function,
\begin{equation}
\begin{split}
\vevB{\cO_0^\alpha(\w)}
&= \frac{1}{4} v^\alpha \langle B_\infty \lvert \bar{b}_v \delta(\img \hat{D}_+ + \w) b_v
\rvert B_\infty \rangle + \ord(\Lambda)
\\
&= \frac{1}{2} v^\alpha f(\w) + \ord(\Lambda)
,\end{split}
\raisetag{4ex}
\end{equation}
where $\lvert B_\infty \rangle $ denotes the $B$ meson state in
the infinite-mass limit and the $b_v$ are the static heavy quark fields.

Together with Eqs.~\eqref{dGqdy} and \eqref{Tc_to-0,0} we find the lepton energy
spectrum at leading order
\begin{subequations}
\label{dGqdy_to-0,0}
\begin{equation}
\frac{\df\Gamma_c}{\df y} = 2 \Gamma_0 y^2 \theta(y)
\int\df\w \theta(\D_\w - \rho) \Gamma_p(\D_\w) f(\w)
,\end{equation}
where ($R = \rho/\D$)
\begin{equation*}
\Gamma_p(\D) = (1 - 3R^2 + 2R^3) + (2 - 3R + R^3)\D
\end{equation*}
contains a purely kinematic $\D$ dependence determined by the parton model.
Letting $\rho \to 0$ we obtain the result for $b \to u$
\begin{equation}
\frac{\df\Gamma_u}{\df y} = 2 \Gamma_0 y^2 \theta(y)
\int\df\w \theta(\D_\w) (1 + 2\D_\w) f(\w)
.\end{equation}
\end{subequations}

Obviously, the leading-order result amounts to convoluting the kinematic
$\D$ dependence of the parton model with $f(\w)$.
The overall factor of $y^2 \theta(y)$ is not convoluted, since it is a
trivial phase space factor, unrelated to the OPE. Thus,
in the modified expansion we keep the factor $y^2 \theta(y)$ and
Eqs.~\eqref{dGqdy_to-0,0} are valid (to $\ord(1)$) over the entire
phase space.

Furthermore, from the first relation in
Eq.~\eqref{hp_lc} it is apparent that the twist expansion
will always yield a convolution of the kinematic $\D$ dependence rather than
the $m_b$ dependence, as originally argued in Ref.~\cite{Mannel:1994pm}.
While a convolution of the $m_b$ dependence is correct to leading order
in the usual twist expansion, it introduces spurious subleading corrections, as has been
noted before. In the modified expansion it already fails at leading order.

\vspace{-1ex}
\subsection{Higher Twist Contributions}
\vspace{-1ex}
The light-cone operators needed to consistently
match all contributions of $\ord(\Lambda)$ and $\ord(\Lambda^2)$ are given by

\begin{widetext}
\begin{subequations}
\label{higher-ops}
\begin{align}
\cO_1^{\alpha\mu}(\w)
&= \iint\df\w_1 \df\w_2 \delta'(\w;\w_1,\w_2)
\bar{b} \delta(\img\hat{\cD}_+ + \w_2) \img\hat{\cD}^\mu \delta(\img\hat{\cD}_+ + \w_1) \gamma^\alpha P_L b
,\\
\cO_2^{\alpha\mu\nu}(\w)
&= \iiint\df\w_1 \df\w_2 \df\bar{\w} \delta''(\w;\w_1,\w_2,\bar{\w})
\bar{b} \delta(\img\hat{\cD}_+ + \w_2) \img\hat{\cD}^{(\mu} \delta(\img\hat{\cD}_+ + \bar{\w}) \img\hat{\cD}^{\nu)} \delta(\img\hat{\cD}_+ + \w_1) \gamma^\alpha P_L b
,\\
\cO_3^{\alpha\mu\nu}(\w)
&= \iint\df\w_1 \df\w_2 \delta'(\w;\w_1,\w_2)
\bar{b} \delta(\img\hat{\cD}_+ + \w_2) \img\hat{\cD}^{(\mu} \img\hat{\cD}^{\nu)} \delta(\img\hat{\cD}_+ + \w_1) \gamma^\alpha P_L b
,\\
\cP_4^{\alpha\mu\nu}(\w)
&= - \frac{g}{2} \iint\df\w_1 \df\w_2 \delta'(\w;\w_1,\w_2)
\bar{b} \delta(\img\hat{\cD}_+ + \w_2) (\varepsilon\cdot\hat{G})^{\mu\nu}
\delta(\img\hat{\cD}_+ + \w_1) \gamma^\alpha P_L b
.\end{align}
\end{subequations}
\end{widetext}%
Here, $(\varepsilon\cdot G)^{\mu\nu} = \varepsilon^{\mu\nu}{}_{\lambda\kappa} G^{\lambda\kappa}$
(with $\varepsilon_{0123} = 1$), and the $\delta$-function factors are
\begin{subequations}
\begin{align}
\delta'(\w;\w_1,\w_2)
&= \frac{\delta(\w - \w_1) - \delta(\w - \w_2)}{\w_1 - \w_2}
,\\
\delta''(\w;\w_1,\w_2,\bar{\w}) &= \frac{\delta'(\w;\w_1,\bar{\w}) - \delta'(\w;\w_2,\bar{\w})}{\w_1 - \w_2}
.\end{align}
\end{subequations}

This operator basis differs from that introduced in Ref.~\cite{BLM1} and
used in previous applications \cite{Bauer:2002yu,Burrell:2003cf} by
the different $\cO_1(\w)$ and the additional operator $\cO_2(\w)$.
We note that this is not an artefact of our modified expansion.
With respect to the usual twist power counting Eq.~\eqref{usual-counting}
both operators are formally of leading order.
Nevertheless, their coefficients are of at
least subleading order in this power counting,
because during the matching procedure one effectively shifts orders from the operators
to their coefficients
by partial integration with respect to $\w$, as will be illustrated later on.
In turn, the $B$ expectation values of $\cO_1(\w)$ and $\cO_2(\w)$ will be
parametrized in terms of derivatives of shape functions. In the final
expression for the spectrum these derivatives are then shifted by partial integration
to act on the OPE coefficients. This ensures
that in the final result the coefficients of all shape functions
(apart from kinematic twist terms) are of usual twist $\ord(1)$.
The same also holds for the contributions of $\cO_3(\w)$
and $\cP_4(\w)$ that are of usual subsubleading twist.

The light-cone OPE of $T_c$ now takes the form
\begin{equation}
\label{Tc_to-1,2}
\begin{split}
T_c
&= \int\df\w \theta(\D_\w - \rho)[
C^{(0,0)}\cdot \cO_0 + C^{(1,1)} \cdot \cO_1
\\ & \quad
+ (C^{(1,2)} + C^{(2,2)} ) \cdot \cO_2 + (D^{(1,2)} + D^{(2,2)} ) \cdot \cO_3
\\ & \quad
+ (E^{(1,2)} + E^{(2,2)} ) \cdot \cP_4 + \ord(\Lambda^3) ](\w)
,\end{split}
\raisetag{3ex}
\end{equation}
where the dots denote the contraction of all Lorentz indices, and
the coefficients are
\begin{subequations}
\begin{equation}
\begin{split}
C^{(0,0)}_\alpha(\w) &= C^{(0)}_\alpha(\w)
= (1 - 3R_\w^2 + 2R_\w^3) n_\alpha
\\ & \quad
+ (2 - 3R_\w + R_\w^3)\D_\w  \bn_\alpha
\\
C'^{(1,1)}_{\alpha\mu}(\w)
&= 2(1 - R_\w^3) (n_\alpha + \D_\w \bn_\alpha) \bn_\mu
\\ & \quad
+ 2(1 - 3 R_\w^2 + 2R_\w^3) \eta_{\perp\alpha\mu}
,\\
C'^{(1,2)}_{\alpha\mu\nu}(\w)
&= 4(1 - R_\w)^3(n_\alpha + \D_\w \bn_\alpha) \eta_{\perp\mu\nu}
,\\
C''^{(2,2)}_{\alpha\mu\nu}(\w)
&= \bigl(2(4 - 3R_\w^2 + 2R_\w^3) n_\alpha
\\ & \quad
+ 6(2 - 2R_\w + R_\w^2) \rho \bn_\alpha \bigr) \bn_\mu \bn_\nu
\\ & \quad
+ 8[(1 - R_\w)^3 + 1 - R_\w^3] \eta_{\perp\alpha(\mu} \bn_{\nu)}
,\end{split}
\end{equation}
and
\begin{equation}
\begin{split}
D^{(1,2)}_{\alpha\mu\nu}(\w)
&= - 3(1 - R_\w)^2 (n_\alpha + \D_\w \bn_\alpha) \eta_{\perp\mu\nu}
,\\
D'^{(2,2)}_{\alpha\mu\nu}(\w)
&= -3(1 - R_\w)\bigl((1 + R_\w) n_\alpha + 2\rho \bn_\alpha \bigr) \bn_\mu \bn_\nu
\\ & \quad
- 6(1 - R_\w)^2 \eta_{\perp\alpha(\mu} \bn_{\nu)}
,\\
E^{(1,2)}_{\alpha\mu\nu}(\w)
&= - 3(1 - R_\w)^2 (n_\alpha + \D_\w \bn_\alpha) n_\mu \bn_\nu/2
,\\
E'^{(2,2)}_{\alpha\mu\nu}(\w)
&= 3(1 - R_\w)^2 \eta_{\perp\alpha[\mu} \bn_{\nu]}
.\end{split}
\raisetag{3ex}
\end{equation}
\end{subequations}
The indices enclosed in round or square brackets are completely
symmetrized or antisymmetrized, respectively.
The first superscript denotes the order of the coefficient's term
in the OPE \eqref{Tc_to-1,2}
as it appears in the usual twist expansion
(i.e., the order of the respective shape function
once the $B$ expectation value is taken),
while the second superscript denotes the order of the
coefficients's term in our modified expansion.

For the above reasons, we only quote the derivatives of the OPE coefficients,
as these are the $\ord(1)$ coefficients which will eventually
enter the energy spectrum. The OPE coefficients are obtained by
integrating over $\w$, which increases their order in the usual twist
power counting. The constants of integration are such that
each integral vanishes at $\D_\w = \rho$, that is, the kinematic
$\theta$-function does not contribute to the partial integrations.

We emphasize that the OPE \eqref{Tc_to-1,2} is valid to $\ord(\Lambda)$ over the entire
phase space
and to $\ord(\Lambda^2)$ away from the endpoint. When
expanded into local operators, it correctly reproduces the full
result to $\ord(\Lambda^2)$ \cite{ManoharWise},
as well as all local $\ord(\Lambda^3)$ terms \cite{Gremm:1996df}
corresponding to leading and subleading order in the usual twist expansion.

\subsection{Remarks on the Matching Procedure}

It is worthwhile to point out a subtlety in the matching procedure
leading to Eq.~\eqref{Tc_to-1,2}.
The $\ord(\Lambda)$ terms contained in Eq.~\eqref{Tc_bb} are
\begin{equation*}
\theta(\D_+ - \rho) C'^{(1,1)}_{\alpha\mu}(-\hk_+) \hk^\mu \bar{u}_b \gamma^\alpha P_L u_b
,\end{equation*}
and can be written as a convolution in two ways,
\begin{equation*}
\begin{split}
C'^{(1,1)}_{\alpha\mu}(-\hk_+) \hk^\mu
&= \int\df\w  C'^{(1,1)}_{\alpha\mu}(\w) \hk^\mu \delta(\hk_+ + \w)
\\
&= \int\df\w C^{(1,1)}_{\alpha\mu}(\w) (-\hk^\mu \delta'(\hk_+ + \w))
,\end{split}
\end{equation*}
corresponding to the two possibilities for the matching
\begin{equation}
\label{extr}
C'^{(1,1)}_{\alpha\mu}(\w) \{\img\hat{\cD}^\mu, \delta(\img\hat{\cD}_+ + \w)\}/2
,\quad
C^{(1,1)}_{\alpha\mu}(\w) \cO_1^{\alpha\mu}(\w)
.\end{equation}
This ambiguity has to be resolved by studying the one-gluon matrix element,
since $\hk^\mu$ commutes with the $\delta(\hk_+ + \w)$ function, while
the covariant derivative and the $\delta(\img\hat{\cD}_+ + \w)$ function do not.
The gluon has momentum $\hl \sim \ord(\Lambda)$, a polarization vector $\e = T^a \e^a$,
and we work in light-cone gauge, $A_+ = 0$. In accordance with Eq.~\eqref{mod-counting},
we treat $\hl_+$ exactly and  expand only in $\hl_-$, $\hl_\perp$, $\hep_-$, and $\hep_\perp$.
To $\ord(\Lambda)$ we find
\begin{widetext}
\begin{equation}
\langle b \lvert T_c \rvert bg \rangle
= -g
\frac{\theta(\D_+ - \rho) C^{(1,1)}_{\alpha\mu}(-\hk_+) -
\theta(\D_+ + \hl_+ - \rho) C^{(1,1)}_{\alpha\mu}(-\hk_+ - \hl_+)}{\hl_+}
\hep^\mu
\bar{u}_b \gamma^\alpha P_L u_b + \cO(\Lambda^2)
,\end{equation}
\end{widetext}
showing that we have to match onto $\cO_1(\w)$. Taking the massless limit
(i.e., $\rho \to 0$, $R_\w \to 0$) we
note that this is
in disagreement with the results of Ref.~\cite{Bauer:2002yu}, where
the first possibility in Eq.~\eqref{extr} has been chosen. Note that the
equations of motion of heavy quark effective theory (HQET)
cannot be used for the operator $\cO_1(\w)$, since the
covariant derivative does not act directly on the heavy quark fields.

A similar problem occurs in the comparison of our $\ord(\Lambda^2)$ contributions with
the ones in Ref.~\cite{Bauer:2002yu}. In our case these contributions are more
complicated, requiring the two different operators $\cO_2(\w)$ and $\cO_3(\w)$.
This is again in contrast with Ref.~\cite{Bauer:2002yu},
where only $\cO_3(\w)$ appears.
In both cases the differences start at $\ord(\Lambda^3)$ in the
local expansion of the operators, and thus also in the final spectrum,
as we will see below.

While this paper was in the review process, studies of
$B \to X_u \ell \bar{\nu}_\ell$  based on
``soft collinear effective theory'' (SCET) appeared which shed some light
on these differences \cite{Bosch:2004cb,Lee:2004ja,Beneke:2004in}.
The SCET-based calculations show that the basis
introduced originally in Ref.~\cite{BLM1} is a complete basis of subleading
operators, at least at tree level. However, in all these cases the light-cone vectors
are defined based on the momentum $m_b v - q$, where $q$ is the
total leptonic momentum. Here and in Ref.~\cite{Bauer:2002yu} a different choice of
light-cone vectors is used, which is based
on $m_b v - p_\ell$, where $p_\ell$ is the momentum of the charged lepton only.
It should be possible to relate the two choices by a coordinate
transformation, i.e., by a reparametrization. We shall not go into any
details here, but our results show that for the latter choice of light-cone coordinates
the operator basis in Ref.~\cite{Bauer:2002yu} is incomplete.

\section{The Lepton Energy Spectrum}

\subsection{Shape Functions}

In the last step we need to parametrize the $B$ expectation values of the
operators \eqref{leading-op} and \eqref{higher-ops}. To be consistent
with our modified expansion we have to include all shape functions of
leading and subleading order in the usual twist power counting, but also
those of usual subsubleading twist with moments of local $\ord(\Lambda^2)$.

The expansion of the QCD fields and states
into HQET ones produces many additional operators [e.g., the $O_1(\w)$ and $P_2(\w)$ of
Refs.~\cite{BLM1,Bauer:2002yu}] and shape functions. However, these higher-order shape functions
always occur in particular combinations with those arising at leading order
in the HQET expansion and can be suitably combined with them.
We therefore take a different approach and directly parametrize the operators in QCD,
which automatically combines
the leading and higher-order HQET shape functions appropriately.

This is in fact similar to
what is used in the context of the local expansion, where for example
the matrix element corresponding to the kinetic energy operator
$\mu_\pi^2$ is also defined using the states of full QCD, and thus this
matrix element
is equal to the kinetic energy matrix element $\lambda_1$ of HQET only
to leading order in the $1/m_b$ expansion.

For the leading operator we have
\begin{align}
\vevB{2\cO_0^\alpha(\w)}
&= F_0(\w) v^\alpha + K_0(\w) (n - v)^\alpha
,\end{align}
which is exact and defines the two QCD shape functions $F_0(\w)$ and $K_0(\w)$.
They may be expanded into the usual  ones of HQET,
\begin{subequations}
\begin{align}
F_0(\w) &= f(\w) + \frac{1}{2} t(\w) + \ord(\Lambda^3) \delta'(\w)
,\\
K_0(\w) &= \w f(\w) + h_1(\w) + \ord(\Lambda^3) \delta'(\w)
.\end{align}
\end{subequations}
Alternatively, we can directly perform their moment expansions and use
HQET to parametrize their moments,
\begin{subequations}
\begin{align}
\begin{split}
F_0(\w)
&= \delta(\w) - \frac{\hla_0}{2} \delta'(\w) - \frac{\hla_1 + \hta_1}{6} \delta''(\w)
\\ & \quad
- \frac{\hrh_1}{18} \delta'''(\w) + \dotsb
,\end{split}
\\
K_0(\w)
&= \frac{\hla_0 - \hrh_0/2}{3} \delta'(\w) + \frac{\hrh_0}{6} \delta''(\w)
+ \dotsb
,\end{align}
\end{subequations}
where we abbreviated
\begin{gather*}
\hta_1 = \hat{\mathcal{T}}_1 + 3\hat{\mathcal{T}}_2
, \quad
\hta_2 = \hat{\mathcal{T}}_3/3 + \hat{\mathcal{T}}_4
,\\
\hla_0 = \hla_1 + \hta_1 + 3(\hla_2 + \hta_2)
, \quad \hrh_0 = \hrh_1 + 3 \hrh_2
,\end{gather*}
and the $\rho_i$ and $\mathcal{T}_i$ are defined in  \cite{Gremm:1996df}.

We note that $F_0(\w)$ and $K_0(\w)$ are defined in QCD without any
reference to the heavy quark limit. Nevertheless, heavy quark symmetry
still tells us that $K_0(\w)$ is suppressed by one power of $\Lambda$ with
respect to $F_0(\w)$. The normalization of $F_0(\w)$ is exact to all orders in
QCD due to $b$-quark number conservation, while all other moments receive
further corrections of local $\ord(\Lambda^4)$ and higher. The leading
contribution to the neglected moments is also of local $\ord(\Lambda^4)$.

For the higher-twist operators we find
\begin{widetext}
\begin{subequations}
\begin{align}
\vevB{2\cO_1^{\alpha\mu}(\w)}
&= - [\w F_0(\w) v^\alpha + \w K_0(\w) (n - v)^\alpha]' (n - v)^\mu
 - [F_1(\w) v^\alpha + K_1(\w) (n - v)^\alpha ]' n^\mu
 - \frac{1}{2} L_1'(\w) \eta^{\perp\alpha\mu}
,\\
\begin{split}
\vevB{2\cO_2^{\alpha\mu\nu}(\w)}
&= - \frac{1}{4} [G_2(\w) v^\alpha + M_2(\w) (n - v)^\alpha]' \eta^{\perp\mu\nu}
+ \frac{1}{2} [\w^2 F_0(\w) v^\alpha + \w^2 K_0(\w) (n - v)^\alpha ]'' (n - v)^\mu (n - v)^\nu
\\ & \quad
+ \frac{1}{2} [\w F_1(\w) v^\alpha + \w K_1(\w) (n - v)^\alpha ]'' 2n^{(\mu} (n - v)^{\nu)}
+ \frac{1}{2} [F_2(\w) v^\alpha + K_2(\w) (n - v)^\alpha ]'' n^\mu n^\nu
\\ & \quad
+ \frac{1}{2} \eta^{\perp\alpha(\mu} [\w L_1(\w) (n - v)^{\nu)} + L_2(\w) n^{\nu)} ]''
\\
&= -\frac{1}{4} G_2'(\w) v^\alpha \eta^{\perp\mu\nu}
+ \frac{1}{2} [\w^2 F_0(\w)]'' v^\alpha (n - v)^\mu (n - v)^\nu + \dotsb
,\end{split}
\\
\vevB{2\cO_3^{\alpha\mu\nu}(\w)}
&= \frac{1}{2} G_3(\w) v^\alpha \eta^{\perp\mu\nu}
- [\w^2 F_0(\w) ]' v^\alpha (n - v)^\mu (n - v)^\nu + \dotsb
,\\
\vevB{2\cP_4^{\alpha\mu\nu}(\w)}
&= [H_4(\w)(n - v)^\alpha + N_4(\w)v^\alpha] 2 v^{[\mu} n^{\nu]}
- \eta^{\perp\alpha[\mu} [R_4(\w) (n - v)^{\nu]} + S_4(\w) n^{\nu]} ]'
.\end{align}
\end{subequations}
\end{widetext}
These relations are again exact and define the respective shape functions.
For the sake of completeness we give the full parametrization of $\cO_2(\w)$.
In its second line we neglected all Lorentz structures whose shape functions
are of higher order and not needed to the order we are working.
The operator $\cO_3(\w)$ obeys a similar parametrization as
$\cO_2(\w)$, but with different higher-order shape functions.

The moment expansions of the relevant shape functions are
\begin{equation}
\label{moment-exps}
\begin{split}
F_1(\w) &= -\frac{\hla_0}{2} \delta(\w) + \ord(\Lambda^4)\delta'(\w) - \frac{\hrh_1}{18} \delta''(\w) + \dotsb
,\\
L_1(\w) &= \frac{2\hla_0 - \hrh_0}{3} \delta(\w) + \frac{\hrh_0}{3} \delta'(\w) + \dotsb
,\\
G_2(\w) &= -\frac{2(\hla_1 + \hta_1)}{3} \delta'(\w) - \frac{2\hrh_1}{9} \delta''(\w) + \dotsb
,\\
G_3(\w) &= -\frac{2(\hla_1 + \hta_1)}{3} \delta'(\w) + \ord(\Lambda^4) \delta''(\w) + \dotsb
,\\
H_4(\w) &= - (\hla_2 + \hta_2) \delta'(\w) + \ord(\Lambda^4) \delta''(\w) + \dotsb
,\\
R_4(\w) &= -2(\hla_2 + \hta_2) \delta(\w) - \hrh_2 \delta'(\w) + \dotsb
,\\
S_4(\w) &= 2(\hla_2 + \hta_2) \delta(\w) + \ord(\Lambda^4) \delta'(\w) + \dotsb
.\end{split}
\raisetag{19ex}
\end{equation}
The corrections to all moments shown, as well as the first neglected moments,
are again of local $\ord(\Lambda^4)$.
All other shape functions are of at least subsubleading twist in the usual
power counting and do not have moments of local $\ord(\Lambda^2)$, for example
\begin{subequations}
\begin{equation}
\begin{split}
K_1(\w) &= \frac{\hrh_0}{6} \delta'(\w) + \ord(\Lambda^4) \delta''(\w) + \dotsb
,\\
N_4(\w) &= \ord(\Lambda^4) \delta''(\w) + \dotsb
,\end{split}
\end{equation}
and
\begin{equation}
\begin{split}
F_2(\w) &= \ord(\Lambda^4) \delta(\w) + \ord(\Lambda^4) \delta'(\w) + \ord(\Lambda^4) \delta''(\w) + \dotsb
,\\
K_2(\w) &= \ord(\Lambda^4) \delta'(\w) + \dotsb
,\\
L_2(\w) &= \frac{\hrh_0}{3} \delta(\w) + \ord(\Lambda^4) \delta'(\w) + \dotsb
,\\
M_2(\w) &= \ord(\Lambda^4) \delta''(\w) + \dotsb
.\end{split}
\raisetag{8ex}
\end{equation}
\end{subequations}

\subsection{The Spectrum}

In order to write the spectrum in a compact way, it is useful
to define
\begin{equation}
\begin{split}
\Gamma_p^\pm(\D) &= \pm (1 - 3R^2 + 2R^3) + (2 - 3R + R^3)\D
,\\
\Gamma_1^\pm(\D) &= 2(1 - R^3) (\pm 1 + \D)
,\\
\Gamma_2^\pm(\D) &= \pm (4 - 3R^2 + 2R^3) + 3(2 - 2R + R^2)\rho
,\\
\Gamma_3^\pm(\D) &= -3(1 - R)[\pm(1 + R) + 2\rho]
.\end{split}
\end{equation}
The lepton energy spectrum now takes the form
\begin{widetext}
\begin{subequations}
\label{dGqdy_to-2,3}
\begin{equation}
\begin{split}
\frac{\df\Gamma_c}{\df y}
&= 2\Gamma_0 y^2 \theta(y) \int\df\w \theta(\D_\w - \rho)
\Big\{\Gamma_p^+(\D_\w) F_0(\w) + \Gamma_p^-(\D_\w) K_0(\w)
+ \Gamma_1^+(\D_\w) [\w F_0(\w) + 2 F_1(\w) ]
\\ & \quad
+ \Gamma_1^-(\D_\w) [\w K_0(\w) + \dotsb ] + \Gamma_2^+(\D_\w) [\w^2 F_0(\w) + \dotsb ]
+ \Gamma_3^+(\D_\w) [\w^2 F_0(\w) + \dotsb ]
\\ & \quad
+ 2(1 - 3 R_\w^2 + 2R_\w^3) L_1(\w) + 2(1 - R_\w)^3(1 + \D_\w) G_2(\w)
\\ & \quad
- 3(1 - R_\w)^2 \big[ (1 + \D_\w) G_3(\w) - (1 - \D_\w) H_4(\w)
- R_4(\w) - 2S_4(\w) \big] + \ord(\Lambda^3) \Bigr\}
.\end{split}
\end{equation}
Here, $\w K_0(\w)$ and $\w^2 F_0(\w)$ are of usual subsubleading
twist, but have a first moment of local $\ord(\Lambda^2)$.
The ellipsis mean that the same coefficient has more shape functions, which
are of the same order in the usual twist power counting, but have only
higher-order moments, e.g., $K_1(\w)$ or $\w F_1(\w)$.
Taking the limit $\rho \to 0$ we obtain the $b \to u$ result
\begin{equation}
\begin{split}
\frac{\df\Gamma_u}{\df y}
&= 2\Gamma_0 y^2 \theta(y) \int\df\w \theta(\D_\w)
\Big\{(1 + 2\D_\w) F_0(\w) + (-1 + 2\D_\w) K_0(\w)
+ 2(1 + \D_\w) [\w F_0(\w) + 2 F_1(\w) ]
\\ & \quad
+ 2(-1 + \D_\w) [\w K_0(\w) + \dotsb ]
+ 4 [\w^2 F_0(\w) + \dotsb ]
- 3 [\w^2 F_0(\w) + \dotsb ]
\\ & \quad
+ 2 L_1(\w) + 2(1 + \D_\w) G_2(\w)
- 3 \bigl[(1 + \D_\w) G_3(\w) - (1 - \D_\w) H_4(\w)
- R_4(\w) - 2S_4(\w)\bigr] + \ord(\Lambda^3) \Bigr\}
.\end{split}
\end{equation}
\end{subequations}

As for the light-cone OPE \eqref{Tc_to-1,2}, Eqs.~\eqref{dGqdy_to-2,3}
are valid to $\ord(\Lambda)$ for all lepton energies
and to $\ord(\Lambda^2)$ away from
the endpoint region. They provide the correct interpolation between
the two regimes of the local expansion and the usual twist expansion.

Employing the moment expansions \eqref{moment-exps} of the various
shape functions, our results reproduce the full result to local $\ord(\Lambda^2)$
\cite{ManoharWise} and all local $\ord(\Lambda^3)$ contributions \cite{Gremm:1996df}
belonging to leading and subleading
order in the usual twist power counting.

To compare our $b \to u$ result with that of Ref.~\cite{Bauer:2002yu}, we
neglect all terms that are of subsubleading twist in the usual twist
power counting and include
the overall $y^2 = (1 - \D)^2$ in the power counting. Also dropping the
$\theta(y)$ in front and expanding $F_0(\w)$ and $K_0(\w)$ into HQET
shape functions, we have
\begin{equation}
\begin{split}
\frac{\df\Gamma_u}{\df y}
&= 2\Gamma_0 \int\df\w \theta(\D_\w) \biggl\{
(1 - \w) f(\w) + \frac{1}{2}t(\w) - h_1(\w)
+ 4F_1(\w) + 2G_2(\w) - 3 G_3(\w) + 3 H_4(\w) \biggr\}
\\
&= \Gamma_0 \biggl\{2\theta(\D) - \frac{\hla_1}{3} [\delta(\D) + \delta'(\D)]
- 11\hla_2 \delta(\D)
- \frac{\hrh_1}{3} \Bigl[5\delta'(\D) + \frac{1}{3}\delta''(\D) \Bigr]
- \hrh_2 \delta'(\D) - \frac{\hta_1}{3} \delta'(\D) \biggr\}
.\end{split}
\end{equation}
\end{widetext}

The second line shows the expansion into local terms.
Both expressions disagree with Ref.~\cite{Bauer:2002yu}.
The differences arise from the new shape functions $F_1(\w)$
and $G_2(\w)$, introduced by $\cO_1(\w)$ and $\cO_2(\w)$, and are explicit
in the coefficient of the $\hrh_1 \delta'(\D)$ term,
which is correctly contained in our results.
In Ref.~\cite{Bauer:2002yu} these new shape functions are
effectively set to $F_1(\w) = 0$ and $G_2(\w) = G_3(\w)$.

\section{Conclusions}

Since $m_c^2 \sim m_b \Lambda_\text{QCD}$ the endpoint
region in $B \to X_c \ell \bar{\nu}_\ell$ is affected by shape-function effects.
In the present paper we have considered these effects to subleading order
in the twist expansion.

The usual twist expansion is valid in the endpoint region only; however,
this expansion can easily be modified to become valid over the full phase
space, thereby yielding a smooth expression for differential rates for any value
of the kinematical variables. We have given the relevant expression for the
spectrum up to $\ord(1/m_b^2)$ and to subleading order in the twist expansion.
Furthermore, in a similar fashion as for the heavy quark expansion parameters,
we suggest to define shape functions using the full field operators and
the full QCD states.

Considering the limiting case $m_c \to 0$ we reveal an inconsistency in
previous work concerning the matching onto subleading shape functions.
It turns out that with our specific choice of light-cone coordinates additional
operators are needed to obtain a complete
set of subleading non-local operators. In this way, the number of functions
needed to describe the subleading twist effects increases.

The results of this paper may be useful for the estimation of higher
moments and for  higher-order corrections to the lower moments
of the lepton spectrum or the hadronic invariant mass spectrum in
$B \to X_c \ell \bar{\nu}_\ell$. Furthermore, since
$B \to X_c \ell \bar{\nu}_\ell$ has a sensitivity to the light-cone
distribution functions
of the $B$ meson one could also make an attempt to extract the shape
functions from this process. However, for a consistent treatment one would
need to include also radiative corrections, which could be considered in the
framework of SCET.

\begin{acknowledgments}
F.T. likes to thank the BaBar group in Dresden for its kind
hospitality while this work was completed. T.M. likes to thank I. Bigi, M. Kraetz, and
N. Uraltsev for discussions related to this subject and acknowledges the
support from the German Ministry for Education and Research (BMBF).
\end{acknowledgments}


\begin{thebibliography}{99}
\bibitem{Benson:2003kp}
D.~Benson, I.~I.~Bigi, T.~Mannel, and N.~Uraltsev,
Nucl.\ Phys.\ B {\bf 665}, 367 (2003), [hep-ph/0302262].
%
\bibitem{Bauer:2004ve}
C.~W.~Bauer, Z.~Ligeti, M.~Luke, A.~V.~Manohar, and M.~Trott,
Phys.\ Rev.\ D\ {\bf 70}, 094017 (2004),
\newline [hep-ph/0408002].
%
\bibitem{Aubert:2004aw}
BABAR Collaboration, B.~Aubert {\it et al.},
Phys.\ Rev.\ Lett.\  {\bf 93}, 011803 (2004),
[hep-ex/0404017].
%
\bibitem{BLM1}
C.~W.~Bauer, M.~E.~Luke, and T.~Mannel,
Phys.\ Rev.\ D {\bf 68},  094001 (2003),
[hep-ph/0102089].
%
\bibitem{Bauer:2002yu}
C.~W.~Bauer, M.~Luke, and T.~Mannel,
Phys.\ Lett.\ B {\bf 543}, 261 (2002),
[hep-ph/0205150].
%
\bibitem{BLM21}
A.~K.~Leibovich, Z.~Ligeti, and M.~B.~Wise,
Phys.\ Lett.\ B {\bf 539},  242 (2002),
[hep-ph/0205148].
%
\bibitem{Burrell:2003cf}
C.~N.~Burrell, M.~E.~Luke, and A.~R.~Williamson,
Phys.\ Rev.\ D {\bf 69}, 074015 (2004),
[hep-ph/0312366].
%
\bibitem{Mannel:1994pm}
T.~Mannel and M.~Neubert,
Phys.\ Rev.\ D {\bf 50}, 2037 (1994),
[hep-ph/9402288].
%
\bibitem{ManoharWise}
A.~V.~Manohar and M.~B.~Wise,
Phys.\ Rev.\ D {\bf 49}, 1310 (1994),
[hep-ph/9308246].
%
\bibitem{Bigiinclusive1}
I.~I.~Y.~Bigi, M.~A.~Shifman, N.~G.~Uraltsev, and A.~I.~Vainshtein,
Phys.\ Rev.\ Lett.\  {\bf 71},  496 (1993),
\newline [hep-ph/9304225].
%
\bibitem{Mannelinc}
T.~Mannel,
Nucl.\ Phys.\ B {\bf 413}, 396 (1994),
\newline [hep-ph/9308262].
%
\bibitem{NeubertShape}
M.~Neubert,
Phys.\ Rev.\ D {\bf 49}, 3392 (1994),
\newline [hep-ph/9311325].
%
\bibitem{NeubertShape1}
M.~Neubert,
Phys.\ Rev.\ D {\bf 49},  4623 (1994),
\newline [hep-ph/9312311].
%
\bibitem{BigiMotion}
I.~I.~Y.~Bigi, M.~A.~Shifman, N.~G.~Uraltsev, and A.~I.~Vainshtein,
Int.\ J.\ Mod.\ Phys.\ A {\bf 9},  2467 (1994),
[hep-ph/9312359].
%
\bibitem{Gremm:1996df}
M.~Gremm and A.~Kapustin,
Phys.\ Rev.\ D {\bf 55}, 6924 (1997),
[hep-ph/9603448].
%
\bibitem{Lee:2004ja}
K.~S.~M.~Lee and I.~W.~Stewart,
hep-ph/0409045.
%
\bibitem{Bosch:2004cb}
S.~W.~Bosch, M.~Neubert, and G.~Paz,
JHEP {\bf 11}, 073 (2004),
[hep-ph/0409115].
%
\bibitem{Beneke:2004in}
M.~Beneke, F.~Campanario, T.~Mannel, and B.~D.~Pecjak,
hep-ph/0411395.

\end{thebibliography}
\end{document}